\documentclass[twocolumn,showpacs,preprintnumbers,amsmath,amssymb]{revtex4}

\newcommand\beq{\begin{equation}}
\newcommand\eeq{\end{equation}}

\usepackage{graphicx}
\usepackage{dcolumn}
\usepackage{bm}

\begin{document}

\preprint{LIGO-P060003-01-R}


\title{On the Analytic Structure of a Family of Hyperboloidal Beams\\
of Potential Interest for Advanced LIGO}

\author{Vincenzo Galdi}\email{vgaldi@unisannio.it} \homepage{http://www.ing.unisannio.it/vgaldi}
\author{Giuseppe Castaldi}%
\author{Vincenzo Pierro}%
\author{Innocenzo M. Pinto}%
\affiliation{
Waves Group, Department of Engineering, University of Sannio, I-82100 Benevento, Italy\\
}%

\author{Juri Agresti}
\author{Erika D'Ambrosio}
\author{Riccardo DeSalvo}
\affiliation{LIGO Laboratory, California Institute of Technology, Pasadena, CA 91125}%
\date{\today}

\begin{abstract}
For the baseline design of the advanced Laser Interferometer Gravitational-wave Observatory (LIGO), use of optical cavities with {\em non-spherical} mirrors supporting flat-top (``mesa'') beams, potentially capable of mitigating the thermal noise of the mirrors, has recently drawn a considerable attention. 
To reduce the severe tilt-instability problems affecting the originally conceived {\em nearly-flat}, ``Mexican-hat-shaped'' mirror configuration, K. S. Thorne proposed a {\em nearly-concentric} mirror configuration capable of producing the same mesa beam profile on the mirror surfaces. Subsequently, Bondarescu and Thorne introduced a generalized construction that leads to a one-parameter family of ``hyperboloidal'' beams which allows continuous spanning from the nearly-flat to the nearly-concentric mesa beam configurations.
This paper is concerned with a study of the analytic structure of the above family of hyperboloidal beams. Capitalizing on certain results from the applied optics literature on flat-top beams, a physically-insightful and computationally-effective representation is derived in terms of rapidly-converging Gauss-Laguerre expansions. Moreover, the functional relation between two {\em generic} hyperboloidal beams is investigated. This leads to a generalization (involving {\em fractional} Fourier transform operators of {\em complex} order) of some recently discovered {\em duality} relations between the nearly-flat and nearly-concentric mesa configurations. Possible implications and perspectives for the advanced LIGO optical cavity design are discussed.
\end{abstract}

\maketitle

\section{Introduction}
\label{Intro}
The current baseline design for the Laser Interferometer Gravitational-wave Observatory (LIGO) \cite{LIGO}, as well as that for its {\em advanced} version \cite{Shoemaker2003}, is based on the use of Fabry-Perot optical cavities composed of {\em spherical} mirrors, which support {\em standard} Gaussian beams (GBs) \cite{Siegman}. During the past few years, there has been a growing interest toward the use of {\em non-spherical} mirrors as a possible aid for reducing the thermal noise of the mirrors \cite{Thorne2000a}. In particular, it was proposed by D'Ambrosio, O'Shaughnessy and Thorne \cite{Thorne2000b} to replace the GB profile with a {\em flat-top} (commonly referred to as ``mesa'') profile, for better averaging the thermally-induced mirror surface fluctuations. They showed that such mesa beams could be synthesized via coherent superposition of minimum-spreading GBs with parallel optical axes, and could be supported by {\em nearly-flat}, ``Mexican-hat-shaped'' mirrors \footnote{This nickname refers to the shape of the mirror, which resembles a ``sombrero'' hat.}. A thorough investigation of the theoretical implications and implementation-related issues \cite{Dambrosio2003b,Dambrosio2003a,Shaugh2004,Dambrosio2004a,Dambrosio2004b} indicated a potential reduction by a factor three in the thermoelastic noise power and a factor two in the coating Brownian thermal noise power, without substantial fabrication impediments. A prototype optical cavity is currently being developed, and experimental tests are under way \cite{Agresti2004}.

In gravitational-wave interferometers, a serious concern is posed by the light-pressure-induced {\em tilt-instability} of the cavity mirrors. In this connection, the inherent tilt-instability of the current (nearly-flat, spherical mirror) LIGO baseline design, was first pointed out by Sidles and Sigg \cite{Sidles2003}. Subsequently, Savov and Vyatchanin \cite{Savov2004} found similar effects for the nearly-flat mesa (FM) designs. Based on these observations, and on the results by Sidles and Sigg \cite{Sidles2003} concerning the comparison between nearly-flat and nearly-concentric spherical mirrors, Thorne proposed an alternative {\em nearly-concentric} Mexican-hat-shaped mirror configuration, capable of supporting mesa beams with intensity distribution at the mirror identical with that of the FM configuration, but featuring a {\em much weaker} \footnote{Actually, even weaker than the nearly-concentric, spherical design (supporting GBs) proposed by Sidles and Sigg \cite{Sidles2003}.} tilt-instability \cite{Savov2004}. These nearly-concentric mesa (CM) beams are synthesized by coherent superposition of minimum-spreading GBs with non-parallel optical axes sharing a common point. 
Quite remarkably, the FM and CM configurations were found to be connected through a {\em duality} relation,
first discovered numerically by Savov and Vyatchanin \cite{Savov2004}, and subsequently proved analytically by Agresti {\em et al.} \cite{Agresti2005a}, which allows a {\em one-to-one} mapping between {\em all} the corresponding eigenmodes. The geometrical construction underlying FM and CM beams was further generalized by Bondarescu and Thorne \cite{Bondarescu2004}, in terms of a family of ``hyperboloidal'' beams, parameterized by a ``twist-angle'' $\alpha\in[0,\pi]$ which allows continuous spanning from the FM ($\alpha=0$) to the CM ($\alpha=\pi$) configurations, passing through the standard GB ($\alpha=\pi/2$) case. It was suggested in \cite{Bondarescu2004} that the optimal configuration, in terms of both thermal-noise and tilt-instability reduction, should be found in a neighborhood of $\alpha=\pi$ (CM configuration). This renders the family of Bondarescu-Thorne (BT) hyperboloidal beams of potential interest in the design of advanced LIGO.

This paper elaborates on the analytic structure of the BT hyperboloidal beams. Our investigation capitalizes on and generalizes a number of results from the applied optics literature concerning flat-top beams, which have most likely not come to the attention of the gravitational wave community. Indeed, during the past decade, flat-top beams have drawn a considerable attention from the applied optics community, and several models have been proposed and investigated. Prominent among them are the celebrated ``supergaussian'' beams \cite{Siegman,Svelto1988}, the ``flattened Gaussian'' beams introduced by Gori and co-workers \cite{Gori1994,Bagini1996,Borghi2001}, the ``flattened'' beams introduced by Sheppard and Saghafi \cite{Sheppard1996}, the  ``flat-topped multi-Gaussian'' beams introduced by Tovar \cite{Tovar2001}, and the ``flat-topped'' beams introduced by Li \cite{Li2002a,Li2002b} \footnote{The reader is referred to \cite{Santarsiero1999,Lu2002} for examples of numerical investigations aimed at ascertaining the correspondence between different models under appropriate parameter conditions.}. With the exception of the analytically-intractable supergaussian, all other models admit {\em analytic} parameterizations in terms of Gaussian \cite{Tovar2001,Li2002a,Li2002b}, Gaussian-Laguerre (GL) \cite{Gori1994,Bagini1996,Sheppard1996}, or ``elegant'' GL \cite{Borghi2001} beam expansions. In this paper, we first show that the FM and CM beams belong to the class of flattened beams introduced in \cite{Sheppard1996}, and can therefore be represented in terms of the {\em rapidly-converging} GL beam expansions derived therein. Based on this observation, we then generalize the approach in \cite{Sheppard1996} to accommodate the more general family of BT hyperboloidal beams \cite{Bondarescu2004}. This leads to a generalization (at least for the dominant eigenmode) of the duality relations discovered in \cite{Agresti2005a}, which involves {\em fractional} Fourier transforms of {\em complex} order. The above results, here discussed for the simplest case of the dominant eigenmode, set the stage for the development of new {\em problem-matched} computational tools for the modal analysis of Fabry-Perot optical cavities supporting {\em general} BT hyperboloidal beams.

The remainder of the paper is laid out as follows. Section \ref{Mesa} introduces the problem geometry, and provides a compact review of background results from \cite{Bondarescu2004} on BT hyperboloidal beams and supporting mirrors. Section \ref{Analytic} contains the analytic derivations concerning the GL expansions and the generalized duality relations, as well as representative numerical results for validation and calibration. Section \ref{Conclusions} contains preliminary conclusions and recommendations.

%
\begin{figure}
\begin{center}
\includegraphics[width=8cm]{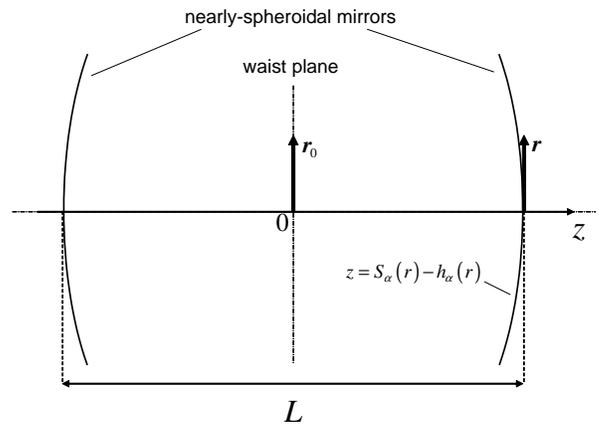}
\end{center}
\caption{Problem schematic: A perfectly symmetric Fabry-Perot optical cavity composed of two nearly-spheroidal mirrors separated by a distance $L$ along the $z$-axis.
The transverse coordinates at the waist ($z=0$) and mirror ($z=L/2$) planes are denoted by ${\bf r}_0$ and ${\bf r}$, respectively. For the $\alpha$-parameterized family of hyperboloidal beams of interest, the mirror shape is obtained by adding the perturbation $-h_{\alpha}$ in (\ref{eq:halpha}) to the fiducial spheroid in (\ref{eq:Salpha}).}
\label{Figure1}
\end{figure}

%
\begin{figure}
\begin{center}
\includegraphics[width=6cm]{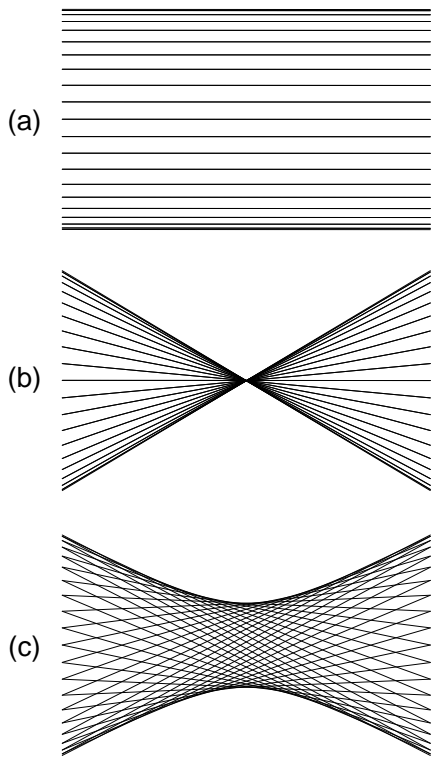}
\end{center}
\caption{Geometrical construction of the BT hyperboloidal beams in (\ref{eq:UBT}): Minimum-spreading GBs are launched from a circular equivalent aperture of radius $R_0$ at the waist plane ($z=0$), with optical axes pointing along the unit vector ${\bf u}_{\alpha}$ in (\ref{eq:ualpha}). (a) FM beam ($\alpha=0$): Optical axes are the generators of a cylinder. (b) CM beams ($\alpha=\pi$): Optical axes are the generators of a cone. (c) Generic hyperboloidal beams ($0\leq \alpha \leq \pi$): Optical axes are the generators of a hyperboloid. }
\label{Figure2}
\end{figure}

\section{Background: BT Hyperboloidal Beams and Supporting Mirrors}
\label{Mesa}
In this Section, we briefly review the procedure proposed by Bondarescu and Thorne \cite{Bondarescu2004} for constructing a family of ``hyperboloidal'' beams, which contains as special limiting cases the FM and CM beams. Referring to the problem geometry illustrated in Fig. \ref{Figure1}, we consider a perfectly symmetric Fabry-Perot optical cavity composed of two nearly-spheroidal mirrors separated by a distance $L$ along the $z$-axis of a Cartesian $(x,y,z)$ (and associated cylindrical $(r,\theta,z)$) coordinate system. The transverse coordinates at the waist ($z=0$) and mirror ($z=L/2$) planes are denoted by ${\bf r}_0\equiv x_0 {\bf \hat x}+ y_0 {\bf \hat y}=r_0\cos\theta_0 {\bf \hat x}+r_0\sin\theta_0 {\bf \hat y}$ and ${\bf r}\equiv x {\bf \hat x}+ y {\bf \hat y}=r\cos\theta {\bf \hat x}+r\sin\theta {\bf \hat y}$, respectively. Here and henceforth, ${\bf \hat x}$, ${\bf \hat y}$ and ${\bf \hat z}$ denote the standard Cartesian unit vectors. Throughout the paper, an implicit time-harmonic $\exp(-i\omega t)$ dependence is assumed for all field quantities.

The BT construction \cite{Bondarescu2004}, which generalizes the original idea in \cite{Thorne2000b}, is based on the superposition of minimum-spreading GBs launched from a circular equivalent aperture of radius $R_0$ at the waist plane ($z=0$), with optical axes pointing along the unit vector  
\begin{eqnarray}
{\bf u}_{\alpha}&=&\frac{r_0}{L}\left[\cos\theta_0-\cos(\theta_0-\alpha)\right]{\bf \hat x}\nonumber\\
&+&\frac{r_0}{L}\left[\sin\theta_0-\sin(\theta_0-\alpha)\right]{\bf \hat y}+{\bf \hat z}.
\label{eq:ualpha}
\end{eqnarray}
As shown in Fig. \ref{Figure2}, these optical axes are the generators of hyperboloids. The ``twist-angle'' $\alpha$ in (\ref{eq:ualpha}) parameterizes this family of ``hyperboloidal'' beams, allowing {\em continuous} spanning from the FM configuration ($\alpha=0$, cylindrical degenerate, cf. Fig. \ref{Figure2}(a)) to the CM configuration ($\alpha=\pi$, conical degenerate, cf. Fig. \ref{Figure2}(b)). For this family of beams, the wavefronts at the mirror location are roughly approximated by the ``fiducial'' spheroids \cite{Bondarescu2004}
\begin{eqnarray}
z&=&S_{\alpha}(r)\equiv
\sqrt{\left(\frac{L}{2}\right)^2-r^2\sin^2(\alpha/2)}\nonumber\\
&\approx&\frac{L}{2}-\frac{r^2\sin^2(\alpha/2)}{L},~~r\ll L/2,
\label{eq:Salpha}
\end{eqnarray}
which degenerate into {\em planar} and {\em spherical} surfaces in the FM ($\alpha=0$) and CM ($\alpha=\pi$) case, respectively. Following \cite{Bondarescu2004}, an integral expression (valid under the paraxial approximation) for the (unnormalized) beam field distribution on these fiducial surfaces can be written as
\begin{eqnarray}
U_{\alpha}(r,S_{\alpha})&=& \Lambda\int_0^{R_0}dr_0\int_0^{2\pi} d\theta_0
r_0\exp\left[
 i \frac{rr_0}{w_0^2}\sin\theta_0\sin\alpha
\right.\nonumber\\
&-& \left.
\frac{\left(r^2+r_0^2-2rr_0\cos\theta_0\right)}{2w_0^2}\left(1-i\cos\alpha\right)\right].
\label{eq:UBT}
\end{eqnarray}
In (\ref{eq:UBT}), $\Lambda$ is an $\alpha$-independent complex constant, and $w_0$ is the GB spot size at the waist. According to the minimum-spreading criterion, this spot size is chosen as
\beq
w_0=\sqrt{\frac{L}{k_0}}, 
\eeq
where $k_0=2\pi/\lambda_0$ denotes the free-space wavenumber ($\lambda_0$ denoting the free-space wavelength), so that the mirror plane is located exactly at the Rayleigh distance \cite{Siegman},
\beq
z_R\equiv \frac{k_0 w_0^2}{2}=\frac{L}{2}.
\label{eq:Rayleigh}
\eeq
Note that the expression in (\ref{eq:UBT}) is valid {\em only} on the fiducial surface $z=S_{\alpha}(r)$. 
For $\alpha=\pi/2$, the double integral in (\ref{eq:UBT}) can be computed in closed form, yielding a simple Gaussian \cite{Bondarescu2004},
\beq
U_{\pi/2}=\Lambda_0 \exp\left(-\frac{r^2}{2w_0^2}\right),
\label{eq:Upi2}
\eeq
with $\Lambda_0$ denoting a complex constant. For other values of $\alpha$, the radial integral in (\ref{eq:UBT}) can still be computed analytically, whereas the angular integral has to be evaluated numerically. It is readily verified from (\ref{eq:UBT}) that the following symmetry relations hold: 
\begin{subequations}
\begin{eqnarray}
U_{-\alpha}=U_{\alpha},\label{eq:symm1}\\
\frac{U_{\pi-\alpha}}{\Lambda}=\frac{U^*_{\alpha}}{\Lambda^*},
\label{eq:symm2}
\end{eqnarray}
\label{eq:symm}
\end{subequations}
where $^*$ denotes complex conjugation; this sets the minimal meaningful range for the twist-angle $\alpha$ to $[0,\pi]$. The relation (\ref{eq:symm2}) can be interpreted in the broader {\em duality} framework detailed in \cite{Savov2004,Agresti2005a} (see Section \ref{Duality}). 

From the theory of graded-phase mirrors \cite{Pare1992,Pare1994}, it is well-known that, in order for an optical cavity to support a stable beam with a given profile as the fundamental eigenmode, its mirror profile has to match the beam wavefront. For the BT hyperboloidal beams in (\ref{eq:UBT}), this can be achieved by applying a correction
\beq
h_{\alpha}(r)=\frac{\arg\left[U_{\alpha}(r,S_{\alpha})\right]-\arg\left[U_{\alpha}(0,S_{\alpha})\right]}{k_0}
\label{eq:halpha}
\eeq
to the fiducial spheroidal shape $S_{\alpha}$ in (\ref{eq:Salpha}), so that
\beq
\arg[U_{\alpha}(r,S_{\alpha}-h_{\alpha})]=\mbox{constant}.
\eeq
For the FM ($\alpha=0$) and CM ($\alpha=\pi$) cases, the correction in (\ref{eq:halpha}) reduces to the Mexican-hat-shaped profile in \cite{Dambrosio2003a,Dambrosio2003b} (see also Fig. \ref{Figure6}(a) below).
Moreover, from (\ref{eq:symm2}), the remarkable result
\beq
h_{\pi-\alpha}(r)=-h_{\alpha}(r)
\label{eq:hpial}
\eeq
follows, which can also be interpreted within the above-mentioned duality framework \cite{Savov2004,Agresti2005a}.

\section{Analytic Structure of BT Hyperboloidal Beams}
\label{Analytic}
Capitalizing on the background results summarized in Section \ref{Mesa}, in this Section, we develop the analytic GL representation of {\em general} BT hyperboloidal beams (valid at {\em any} point in space, within the limit of the paraxial approximation) as well as of the supporting mirror profiles. Moreover, we generalize the symmetry/duality relations in (\ref{eq:symm}) and (\ref{eq:hpial}) to the most general case involving two arbitrary values of the twist-angle parameter. For such generalization, we provide functional and optical interpretations, based on {\em fractional} Fourier operators of {\em complex} order.

\subsection{Field Distribution at Waist}
We begin by considering the field distribution at the waist plane ($z=0$) for the FM ($\alpha=0$) beam \cite{Dambrosio2003a,Agresti2005a},
\begin{eqnarray}
\!\!\!\!\!\!\!\!U_{0}(r,0)&=&\frac{1}{\pi R_0^2}\iint_{r_0\le R_0}d{\bf r}_0 \exp\left(
-\frac{|{\bf r}-{\bf r}_0|^2}{w_0^2}
\right)\nonumber\\
&=&\!\!
\frac{2}{R_0^2}\!\int_{0}^{R_0}\!\!\!\!\!dr_0 r_0 
I_0\!\!\left(\frac{2rr_0}{w_0^2}\!\right)\!
\exp\!\!\left[\!-\frac{(r^2\!+\!r_0^2)}{w_0^2}\right]\!\!,
\label{eq:U0}
\end{eqnarray}
where $I_0(\xi)$ denotes a zeroth-order modified Bessel function of the first kind \cite[Sec. 9.6]{Abramowitz}.

\subsubsection{Duality Relations}
\label{Duality}
In \cite{Agresti2005a}, within a broader framework of duality relations, the CM ($\alpha=\pi$) beam field distribution at the waist plane was shown to be related to the FM one in (\ref{eq:U0}) via a Fourier transform operator
\beq
U_{\pi}(r,0)
\stackrel{{\cal H}_{w_0}}{\longleftrightarrow}
U_{0}(r,0)
\label{eq:HT}
\eeq
which, in view of the assumed cylindrical rotational symmetry, takes the form of the Hankel transform (HT)
\beq
{\cal H}_{w_0}\left[F(r)\right]\equiv
\frac{2}{w_0^2}
\int_{0}^{\infty}dr_0 r_0 F(r_0) J_0\left(\frac{2rr_0}{w_0^2}\right).
\label{eq:Hw0}
\eeq
In (\ref{eq:Hw0}) and henceforth, $J_n(\xi)$ denotes an $n$th-order Bessel function of the first kind \cite[Sec. 9.1]{Abramowitz}. Straightforward application of the HT (\ref{eq:Hw0}) to (\ref{eq:U0}) yields, via the convolution theorem \cite{Agresti2005a},
\begin{eqnarray}
U_{\pi}(r,0)&=&{\cal H}_{w_0}\left[U_{0}(r,0)\right]\nonumber\\
&=&\frac{w_0^2}{rR_0}J_1\left(\frac{2rR_0}{w_0^2}\right)\exp\left(-\frac{r^2}{w_0^2}
\right).
\label{eq:Upi}
\end{eqnarray}

\subsubsection{GL Representation}
\begin{figure}
\begin{center}
\includegraphics[width=8cm]{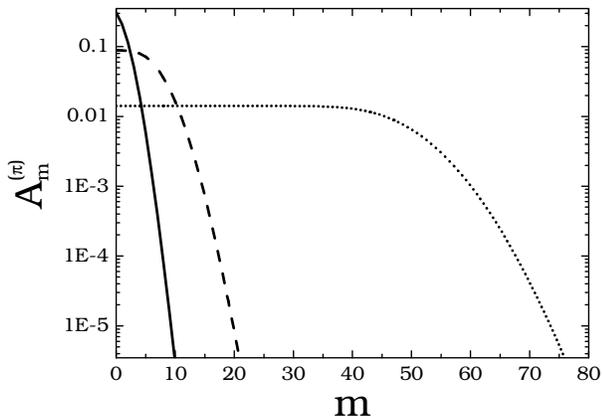}
\end{center}
\caption{GL expansion coefficients $A_m^{(\pi)}$ in (\ref{eq:Apim}) vs. summation index $m$, for different values of $w_0/R_0$. Continuous curve: $w_0/R_0=0.5$; ~~Dashed curve: $w_0/R_0=0.25$; ~~Dotted curve: $w_0/R_0=0.1$.}
\label{Figure3}
\end{figure}

The CM field distribution in (\ref{eq:Upi}) is recognized to coincide with the one used in \cite[Sec. 2]{Sheppard1996} to generate beams with a ``flattened'' far-field profile, for which a GL beam expansion was subsequently derived. Following \cite{Sheppard1996}, the field distribution in (\ref{eq:Upi}) can be expanded as
\beq
U_{\pi}(r,0)=\sum_{m=0}^{\infty}A^{(\pi)}_m \psi_m\left(\frac{\sqrt{2}r}{w_0}\right).
\label{eq:UpiGL}
\eeq
In (\ref{eq:UpiGL}), $\psi_m(\xi)$ are orthonormal GL basis functions,
\begin{subequations}
\beq
\psi_m(\xi)=\sqrt{2}\exp\left(-\frac{\xi^2}{2}\right) L_m(\xi^2),
\label{eq:psim}
\eeq
\beq
\int_0^{\infty}\psi_p(\xi) \psi_q(\xi) \xi d\xi=\delta_{pq},
\eeq
\label{eq:ort}
\end{subequations}
where $L_n(\zeta)$ denotes an $n$th-order Laguerre polynomial \cite[Chap. 22]{Abramowitz}, and $\delta_{pq}$ denotes the Kronecker symbol. The expansion coefficients $A^{(\pi)}_m$ in (\ref{eq:UpiGL}) are given by \cite{Sheppard1996}
\beq
A^{(\pi)}_m=\frac{\sqrt{2}w_0^2}{R_0^2}P\left(m+1,\frac{R_0^2}{2w_0^2}\right),
\label{eq:Apim}
\eeq
where $P(n,\xi)$ denotes an incomplete Gamma function \cite[Eq. (6.5.13)]{Abramowitz}.
The behavior of the expansion coefficients in (\ref{eq:Apim}), as a function of the summation index $m$, is shown in Fig. \ref{Figure3}, for three representative values of the ratio $w_0/R_0$: The $A^{(\pi)}_m$ are almost constant for $m\lesssim R_0^2/(2w_0^2)$, and fall off quite abruptly for $m\gtrsim R_0^2/(2w_0^2)$ \footnote{Quite amusingly, as observed in \cite{Sheppard1996}, they exhibit the same functional form as the spatial behavior of the flattened beams in \cite{Gori1994}.}. For the parametric range of interest for the LIGO design ($w_0/R_0=0.25$), this results in a rapidly-converging ($m<20$) expansion (\ref{eq:UpiGL}). 

The corresponding GL expansion for the FM field distribution at the waist plane can be obtained by exploiting the HT relation in (\ref{eq:HT}). Similar arguments were likewise invoked in \cite[Sec. 4]{Sheppard1996}, in the form of near-to-far-field transformations, to generate beams that were flattened at the waist plane (``inverted flattened'' beams, in the notation of \cite{Sheppard1996}). In this framework, one first observes that the GL basis functions in (\ref{eq:psim}) are eigenfunctions of the HT operator in (\ref{eq:Hw0}) \cite{Yu1998},
\beq
{\cal H}_{w_0}\left[\psi_m\left(\frac{\sqrt{2}r}{w_0}\right)\right]=(-1)^m\psi_m\left(\frac{\sqrt{2}r}{w_0}\right).
\label{eq:eigen}
\eeq
Equation (\ref{eq:eigen}) can be derived as a special case (corresponding, e.g., to letting $a=1$, $b=2$, $\xi=2r/w_0$, $\zeta=r_0/w_0$) of the identity \cite[p. 43]{Erdelyi}  
\begin{eqnarray}
\!\!\!\!\!\!\!\int_0^{\infty} \!\!\!\!&\zeta&\!\!\!
\exp(-a\zeta^2)
L_m(b\zeta^2)
J_0(\xi\zeta)d\zeta
=\frac{(a-b)^m}{2a^{m+1}}\nonumber\\
&\times&\!\!\!\exp\left(-\frac{\xi^2}{4a}\right)
 L_m\left[\frac{b\xi^2}{4a(b-a)}\right],~\mbox{Re}(a)\ge0.
\label{eq:intransf}
\end{eqnarray}
Application of the HT (\ref{eq:Hw0}) to (\ref{eq:UpiGL}) then yields, via (\ref{eq:eigen}), the corresponding GL expansion for the FM field distribution in (\ref{eq:U0}), with the following mapping between the expansion coefficients:
\beq
A^{(0)}_m=(-1)^m A^{(\pi)}_m.
\label{eq:map1}
\eeq
Alternatively, as shown in Appendix \ref{appendix}, one could derive {\em directly} the FM expansion coefficients $A^{(0)}_m$, and then exploit the mapping in (\ref{eq:map1}) to obtain those pertaining to the CM configuration.
The above derivations clarify the relationship between the CM and FM beams in \cite{Bondarescu2004,Agresti2005a} and the GL expansions for the ``flattened'' and ``inverted flattened'' beams in \cite{Sheppard1996}, respectively. 
A natural question then arises, as to whether the expansion coefficient mapping in (\ref{eq:map1}) can be generalized to {\em arbitrary} values of the twist-angle $\alpha$, thereby allowing a GL representation for {\em general} BT hyperboloidal beams.
Recalling that, from (\ref{eq:Upi2}),
\beq
A^{(\pi/2)}_m=0,~~m>0,
\eeq
one notes that the mapping
\beq
A^{(\alpha)}_m=(-\cos\alpha)^m A^{(\pi)}_m
\eeq
accounts correctly for the three notable cases $\alpha=0$ (FM), $\alpha=\pi/2$ (GB) and $\alpha=\pi$ (CM).
One is accordingly led to speculate whether the GL expansion
\beq
U_{\alpha}(r,0)=\sum_{m=0}^{\infty}A^{(\alpha)}_m \psi_m\left(\frac{\sqrt{2}r}{w_0}\right)
\label{eq:GL1}
\eeq
may hold for {\em arbitrary} values of the twist angle $\alpha$. 
This turns out to be indeed the case, as checked by numerical comparison against the BT reference solution in (\ref{eq:UBT}) (see Section \ref{Results} below). The analytic GL expansion in (\ref{eq:GL1}) is obtained here for the first time, to the best of our knowledge, and sets the stage for a generalization, to {\em arbitrary} values of the twist-angle, of the duality relation in (\ref{eq:HT}), whose
possible interpretations and implications are discussed below.

%

\subsubsection{Generalized Duality Relations: Functional and Optical Interpretations}
\begin{figure}
\begin{center}
\includegraphics[width=8cm]{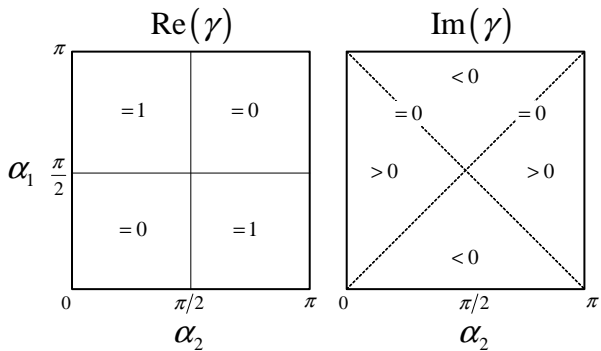}
\end{center}
\caption{Admissible values of the complex order $\gamma$ (\ref{eq:co}) of the generalized HT operator in (\ref{eq:genH0ab}) relating the field distributions of two BT hyperboloidal beams with generic twist-angles, $\alpha_1$ and $\alpha_2$ (cf. (\ref{eq:HTgen})). The real part of $\gamma$ can either be $1$ or $0$, whereas the imaginary part is generally nonzero, except along the two diagonals $\alpha_1=\alpha_2$ ($\sigma=-1$, $\gamma=0$) and $\alpha_1=\pi-\alpha_2$ ($\sigma=1$, $\gamma=1$), where the generalized HT operator in (\ref{eq:genH0ab}) reduces to the identity and ordinary HT operator, respectively.}
\label{Figure4}
\end{figure}

We begin by considering a class of $\sigma$-parameterized modified HT operators
defined as
\begin{subequations}
\begin{eqnarray}
{\cal H}^{(\sigma)}_{w_0}\left[F\left(r\right)\right]
\!\!\!&\equiv&\!\!\!
\frac{4}{w_0^2\left(1+\sigma\right)}
\int_{0}^{\infty}\!\!\!\!r_0 dr_0
F(r_0)J_0\!\!\left[
\frac{4rr_0\sqrt{\sigma}}{w_0^2\left(1+\sigma\right)}
\right]\nonumber\\
\!\!\!&\times&\!\!\!\!
\exp\!\!\left[
\!-\frac{(r^2\!+\!r_0^2)(1\!-\!\sigma)}{w_0^2(1+\sigma)}
\right]\!\!,~~\sigma\ge-1.
\label{eq:genH0}
\end{eqnarray}
For $\sigma<-1$, the integral in (\ref{eq:genH0}) diverges for the beams of interest here (decaying as $O[\exp(-r^2/w_0^2)]$ in the waist plane), and the following definition should be used:
\begin{eqnarray}
\!\!\!\!\!\!\!\!
{\cal H}^{(\sigma)}_{w_0}\left[F\left(r\right)\right]\!\!&\equiv&\!\!
{\cal H}^{(-\sigma)}_{w_0}\left\{
{\cal H}^{(1)}_{w_0}\left[F\left(r\right)\right]
\right\}\nonumber\\
\!\!&=&\!\!{\cal H}^{(1)}_{w_0}\left\{
{\cal H}^{(-\sigma)}_{w_0}\left[F\left(r\right)\right]\right\}\!\!,~\sigma<-1.
\end{eqnarray}
\label{eq:genH0ab}
\end{subequations}
The operator in (\ref{eq:genH0ab}) generalizes the ordinary HT in (\ref{eq:Hw0}); it is readily verified that it reduces to the ordinary HT for $\sigma=1$, and to the identity operator for $\sigma=-1$.
From (\ref{eq:intransf}), it then follows \footnote{Choosing $a=1$, $b=1+\sigma$, $\xi=2\sqrt{2\sigma} r/(w_0\sqrt{1+\sigma})$, $\zeta=\sqrt{2}r_0/(w_0\sqrt{1+\sigma})$.} the generalization of the eigenproblem in (\ref{eq:eigen}), 
\begin{eqnarray}
{\cal H}^{(\sigma)}_{w_0}\left[\psi_m\left(\frac{\sqrt{2}r}{w_0}\right)\right]
=(-\sigma)^m \psi_m\left(\frac{\sqrt{2}r}{w_0}\right).
\label{eq:geneigen}
\end{eqnarray}
Application of the generalized HT (\ref{eq:genH0ab}) to the GL expansion in (\ref{eq:GL1}) reveals, via (\ref{eq:geneigen}), the functional relation between the field distributions at the waist plane pertaining to two BT hyperboloidal beams characterized by {\em generic} values, $\alpha_1$ and $\alpha_2$, of the twist-angle,
\beq
U_{\alpha_2}(r,0)
\stackrel{{\cal H}^{(\sigma)}_{w_0}}{\longleftrightarrow}
U_{\alpha_1}(r,0),~~~\sigma=-\frac{\cos\alpha_2}{\cos\alpha_1}.
\label{eq:HTgen}
\eeq
The generalized HT in (\ref{eq:HTgen}) extends the duality relation in (\ref{eq:HT}) to the most general case, and admits a suggestive analytic interpretation in terms of (the cylindrical version of) a {\em fractional} Fourier operator of {\em complex} order \cite{Wang2002a, Wang2002b, Shih1994, Bernardo1996}. From the physical point of view, complex-order Fourier transform operators can be interpreted in terms of propagation through a paraxial optical system described \cite{Wang2002a, Wang2002b, Shih1994, Bernardo1996} by a {\em complex} ABCD matrix \footnote{The reader is referred to \cite{Siegman,Wang2002a, Wang2002b, Shih1994, Bernardo1996} for alternative optical interpretations in terms of propagation through {\em complex} Gaussian ducts, as well as through optical systems composed of self-imaging components and Gaussian apertures.}. For (\ref{eq:genH0ab}), the ABCD matrix can be shown to be
\begin{subequations}
\begin{eqnarray}
\left[
\begin{matrix}
A & B \cr
C & D
\end{matrix}
\right]&=&\left[
\begin{matrix}
\displaystyle{i\frac{(1-\sigma)}{2\sqrt{\sigma}}} & \displaystyle{\frac{k_0w_0^2(1+\sigma)}{4\sqrt{\sigma}}} \cr
-\displaystyle{\frac{(1+\sigma)}{k_0w_0^2\sqrt{\sigma}}} & \displaystyle{i\frac{(1-\sigma)}{2\sqrt{\sigma}}}
\end{matrix}
\right]\label{eq:ABCD1}\\
&=&\left[
\begin{matrix}
\cos\left(\displaystyle{\frac{\pi\gamma}{2}}\right) & \displaystyle{\frac{L}{2}\sin\left(\frac{\pi\gamma}{2}\right)} \cr
-\displaystyle{\frac{2}{L} \sin\left(\frac{\pi\gamma}{2}\right)} & \cos\left(\displaystyle{\frac{\pi\gamma}{2}}\right)
\end{matrix}
\right].
\label{eq:ABCD2}
\end{eqnarray}
\label{eq:ABCD}
\end{subequations}
In (\ref{eq:ABCD2}), the parameter $\gamma$ denotes the {\em complex} order of the transform, defined as \cite{Wang2002a}
\beq
\gamma\equiv-i\frac{2}{\pi}\log\left[
A\sqrt{\frac{D}{A}}+iB\sqrt{\frac{-C}{B}}
\right]=
1+i\frac{\log(\sigma)}{\pi}.
\label{eq:co}
\eeq
The admissible values of the complex order in (\ref{eq:co}) are illustrated schematically in Fig. \ref{Figure4}, over the meaningful ($\alpha_1$, $\alpha_2$)-range. It is observed that the real part can either be $1$ or $0$, whereas the imaginary part is generally nonzero, except along the two diagonals $\alpha_1=\alpha_2$ and $\alpha_1=\pi-\alpha_2$ where the generalized HT operator in (\ref{eq:genH0ab}) reduces to the identity and ordinary HT operator, respectively.
 
We are currently exploring the possibility of extending the above results, limited to the dominant eigenmode, to higher-order (isotropic and nonisotropic) modes. If successful, such an extension would provide valuable physical insights and computational savings, unraveling the effects of the twist-angle parameter in the eigenspectrum (see also the discussion in Section \ref{Conclusions}). 

%

\subsection{Paraxial Field Distribution: GL Beam Representation}
From the GL expansion at the waist plane in (\ref{eq:GL1}), a paraxially-approximated expression for the field distribution of a generic BT hyperboloidal beam at any point in space can readily be obtained as
\beq
U_{\alpha}(r,z)=\sum_{m=0}^{\infty}A^{(\alpha)}_m \Psi_m\left(r,z\right),
\label{eq:GLBE}
\eeq
where $\Psi_m\left(r,z\right)$ denote the standard GL beam propagators \cite{Siegman}
\begin{eqnarray}
\Psi_m(r,z)&=&\frac{w_0}{w(z)}\psi_m\left[\frac{\sqrt{2}r}{w(z)}\right]
\exp\left[
i\frac{k_0 r^2}{2R(z)}
\right]\nonumber\\
&\times&\exp\left\{
i\left[k_0z-(2m+1)\Phi(z)
\right]
\right\},
\label{eq:GLP}
\end{eqnarray}
which match the GL basis functions at the waist plane, $\Psi_m(r,0)=\psi_m(\sqrt{2}r/w_0)$. In (\ref{eq:GLP}), $w(z)$, $R(z)$ and $\Phi(z)$ denote the standard GB spot size, wavefront radius of curvature, and Gouy phase, respectively \cite{Siegman}
\begin{eqnarray}
w(z)=w_0\sqrt{1+\left(\frac{z}{z_R}\right)^2},~~R(z)=z+\frac{z_R^2}{z},\nonumber\\
\Phi(z)=\arctan\left(\frac{z}{z_R}\right),
\label{eq:GBPAR}
\end{eqnarray}
with the Rayleigh distance $z_R$ defined in (\ref{eq:Rayleigh}). The GL beam expansion in (\ref{eq:GLBE}) represents, together with the generalized duality relation in (\ref{eq:HTgen}), the main original result in this paper.
We stress that, unlike the expression in (\ref{eq:UBT}), the representation in (\ref{eq:GLBE}) is valid {\em at any point} in space, within the limits of the paraxial approximation. When evaluated on the fiducial surface $z=S_{\alpha}(r)$, it yields
\begin{eqnarray}
U_{\alpha}(r,S_{\alpha})&\approx&\frac{(1-i)}{2}
\exp\left[ik_0\left(\frac{L}{2}+\frac{r^2\cos\alpha}{2L}\right)\right]\nonumber\\
&\times&\sum_{m=0}^{\infty}(-i)^m A^{(\alpha)}_m \psi_m\left(\frac{r}{w_0}\right),
\label{eq:USalpha}
\end{eqnarray}
where the approximation $S_{\alpha}\approx L/2$ has been used in (\ref{eq:GBPAR}). It is easily verified that (\ref{eq:USalpha}) satisfies the phase-conjugation symmetry relation in (\ref{eq:symm2}). 
Moreover, by comparing (\ref{eq:USalpha}) to (\ref{eq:GL1}), and recalling (\ref{eq:HTgen}), one can easily derive a generalized duality relation for the field distributions on the fiducial surfaces, in terms of the generalized (complex-order) HT in (\ref{eq:genH0ab}),
\begin{eqnarray}
\!\!\!\!\frac{U_{\alpha_2}(r,S_{\alpha_2})}
{\displaystyle{\exp\left(i\frac{k_0 r^2\!\cos\alpha_2}{2L}\right)}}
\stackrel{{\cal H}^{(\sigma)}_{\sqrt{2}w_0}}{\longleftrightarrow}
\frac{
U_{\alpha_1}(r,S_{\alpha_1})}{\displaystyle{\exp\left(i\frac{k_0 r^2\!\cos\alpha_1}{2L}\right)}},\nonumber\\
\!\!\!\!\!\!\!\sigma=-\frac{\cos\alpha_2}{\cos\alpha_1}.
\label{eq:GENR}
\end{eqnarray}
The relation in (\ref{eq:GENR}) (which is similar to that in (\ref{eq:HTgen}), apart for the phase factors and a scaling by a factor $\sqrt2$ in the GB spot size) generalizes completely the symmetry relations in (\ref{eq:symm}). 

\subsection{Mirror Profile}
The correction to be applied to the fiducial spheroidal mirror shape $S_{\alpha}$ in (\ref{eq:Salpha}) is finally obtained by substituting (\ref{eq:USalpha}) into (\ref{eq:halpha}),
\begin{eqnarray}
h_{\alpha}(r)
\!\!\!&\approx&\!\!\!\!
\frac{1}{k_0}\arg\!\!\left[
\exp\!\!\left(\!
i\frac{k_0 r^2\!\cos\alpha}{2L}
\!\right)
\!\!\frac{\displaystyle{\sum_{m=0}^{\infty}\!(\!-i)^m\!A_m^{(\alpha)}\psi_m\!\!\left(\!\frac{r}{w_0}\!\right)}}{\displaystyle{\sum_{m=0}^{\infty}(-i)^mA_m^{(\alpha)}}\!\!}
\right]\!\!.\nonumber\\
\label{eq:halpha1}
\end{eqnarray}
It is readily verified that, in view of (\ref{eq:eigen}), the correction profile in (\ref{eq:halpha1}) satisfies the duality relation in (\ref{eq:hpial}). Truncation of the infinite series in (\ref{eq:halpha1}) is not an issue, as further discussed below.

\subsection{Representative Results}
\label{Results}
In order to validate and calibrate the proposed GL representations, we now move on to illustrating some representative numerical results. In all examples below, all the relevant parameters were chosen as in \cite{Dambrosio2003a,Bondarescu2004}. More specifically: $L=4$km (length of the optical cavity, cf. Fig. \ref{Figure1}), $\lambda_0=1064$nm (wavelength of the laser beam), $w_0=\sqrt{L\lambda_0/(2\pi)}=2.603$cm (GB spot size at waist), $R_0=4w_0=10.4$cm (radius of the equivalent aperture distribution at the waist plane). For the truncation of the GL series involved, a simple criterion was utilized, requiring that the magnitude of the last retained $M$-th term is less than $0.1\%$ of that of the leading term,
\beq
\left|
\frac{A_M^{(\alpha)}}{A_0^{(\alpha)}}
\right|<10^{-3}.
\label{eq:trunc}
\eeq
For the cases $\alpha=0,\pi$ (see the dashed curve in Fig. \ref{Figure3}), this yields $M=18$. In view of the coefficient mapping in (\ref{eq:map1}), the convergence becomes faster as $\alpha$ approaches the critical value of  $\pi/2$ (pure GB, for which one obtains only one nonzero coefficient).

As a reference solution, we considered the BT integral representation in (\ref{eq:UBT}), where the radial integral was computed analytically, and the angular integration was performed numerically utilizing the adaptive quadrature routines of Mathematica$^{\mbox{\textregistered}}$ \cite{Wolfram}.

\begin{figure}
\begin{center}
\includegraphics[width=8cm]{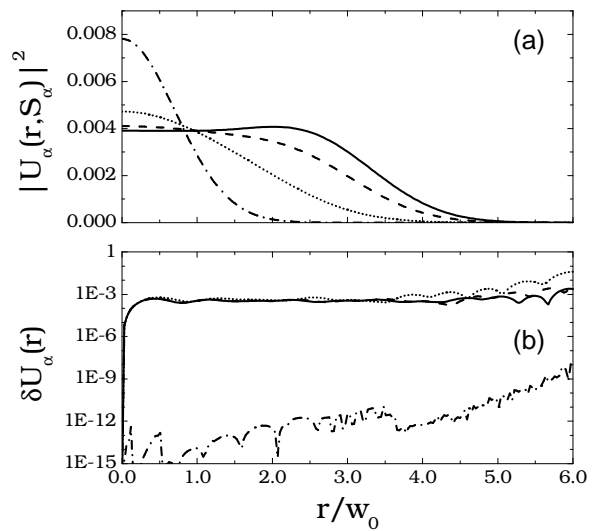}
\end{center}
\caption{BT hyperboloidal beam field distribution evaluated on the fiducial surface $z=S_{\alpha}(r)$, for different values of the twist-angle parameter $\alpha$. Optical cavity parameters: $L=4$km, $\lambda_0=1064$nm, $w_0=\sqrt{L\lambda_0/(2\pi)}=2.603$cm , and $R_0=4w_0=10.4$cm. (a) Intensity distribution computed via (\ref{eq:USalpha}), using the truncation criterion in (\ref{eq:trunc}). (b) Relative error in (\ref{eq:deltaU}). 
Continuous curve: $\alpha=0, \pi (M=18)$; ~~Dashed curve $\alpha=0.1\pi, 0.9\pi (M=17)$; ~~Dotted curve: $\alpha=0.2\pi, 0.8\pi (M=14)$;~~Dotted-dashed curve: $\alpha=0.5\pi (M=0)$.}
\label{Figure5}
\end{figure}

Some representative results for the field distribution are shown in Fig. \ref{Figure5}. Specifically, Fig. \ref{Figure5}(a) shows the GL-computed (via (\ref{eq:USalpha})) intensity distribution on the fiducial surface, for various values of the twist-angle $\alpha$, illustrating the gradual transition from Gaussian ($\alpha=\pi/2$) to
mesa ($\alpha=0,\pi$) profile. To quantify the agreement with the reference solution, Fig. \ref{Figure5}(b) shows the relative error
\beq
\delta U_{\alpha}(r)=\left|
\frac{U_{\alpha}(r)-U^{(BT)}_{\alpha}(r)}{U^{(BT)}_{\alpha}(r)}
\right|,
\label{eq:deltaU}
\eeq
where the superfix $^{(BT)}$ denotes the BT representation in (\ref{eq:UBT}) (where the complex constant $\Lambda$ is determined by enforcing the matching with the GL expansion at $r=0$). 
The error (\ref{eq:deltaU}) never exceeds 0.1\%, over the region of significant field intensity (and drops below numerical precision for the $\alpha=\pi/2$ pure GB case). This is consistent with the truncation criterion in (\ref{eq:trunc}), which can therefore be used to control the accuracy.

\begin{figure}
\begin{center}
\includegraphics[width=8cm]{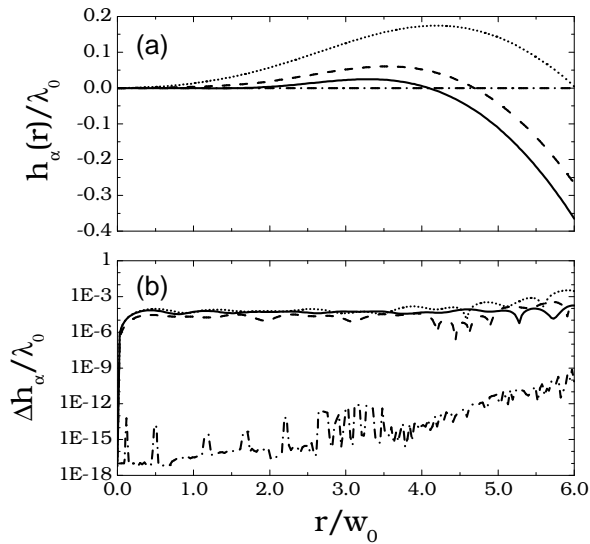}
\end{center}
\caption{Nearly-spheroidal mirror profiles supporting the BT hyperboloidal beams in Fig. \ref{Figure5}. (a) Correction $h_{\alpha}$ to the fiducial spheroid $S_{\alpha}$ in (\ref{eq:Salpha}), computed via (\ref{eq:halpha1}), using the truncation criterion in (\ref{eq:trunc}). (b) Absolute error in (\ref{eq:deltah}).
Continuous curve: $\alpha=\pi (M=18)$; ~~Dashed curve: $\alpha=0.9\pi (M=17)$; ~~Dotted curve: $\alpha=0.8\pi (M=14)$;~~Dotted-dashed curve: $\alpha=0.5\pi (M=0)$. The correction profiles pertaining to $\alpha=0, 0.1\pi, 0.2\pi$ (not shown) differ merely by sign from those pertaining to $\alpha=\pi, 0.9\pi, 0.8\pi$, respectively (cf. (\ref{eq:hpial})).}
\label{Figure6}
\end{figure}

The results pertaining to the mirror profiles are shown Fig. \ref{Figure6}. Specifically, Fig. \ref{Figure6}(a) shows the corrections $h_{\alpha}$ computed via (\ref{eq:halpha1}), illustrating the gradual transition from the spherical ($\alpha=\pi/2$) to the Mexican-hat ($\alpha=0,\pi$) mirror profile. Figure \ref{Figure6}(b) shows the absolute error
\beq
\Delta h_{\alpha}(r)=\left|
h_{\alpha}(r)-h^{(BT)}_{\alpha}(r)
\right|
\label{eq:deltah}
\eeq
with respect to the reference solution. The error (\ref{eq:deltah}) never exceeds $10^{-4}\lambda_0$ over the significantly illuminated portion of the mirror. For the LIGO design ($\lambda_0=1064$nm), this corresponds to errors $\sim$0.1nm, well within the typical fabrication tolerances.

%

%

\section{Conclusions and Recommendations}
\label{Conclusions}
In this paper, the analytic structure of a family of hyperboloidal beams, introduced by Bondarescu and Thorne \cite{Bondarescu2004} as a generalization of the mesa beams supported by Mexican-hat-shaped mirrors, has been investigated. 
Rapidly converging expansions in terms of GL beams  have been first introduced for the ``extremal'' cases of FM and CM beams, capitalizing on results from \cite{Sheppard1996}. The representation has been then extended to the more general BT hyperboloidal beams in \cite{Bondarescu2004}, leading to a complete generalization (for the dominant eigenmode) of the duality relations introduced in \cite{Agresti2005a}, based on {\em fractional} Fourier transforms of {\em complex} order. The above results, numerically validated and calibrated against a reference solution independently-generated from \cite{Bondarescu2004}, provide a physically-insightful and computationally-effective parameterization of the beam and mirror profiles.
It is hoped that they may help addressing the optimization of the advanced LIGO optical cavities in a broader perspective.
In this framework, current and future research directions include:

\begin{itemize}

\item[{\em i)}]{Thorough parametric analysis of the family of BT hyperboloidal beams, as well as of other classes of flat-top beams \cite{Siegman,Svelto1988,Gori1994,Bagini1996,Borghi2001,Tovar2001,Li2002a,Li2002b}, 
aimed at finding optimal design criteria in terms of thermal-noise and tilt-instability reduction.}

\item[{\em ii)}]{Development of {\em semi-analytic, problem-matched} techniques for the computation of higher-order eigenmodes in nearly-spheroidal-mirror optical cavities supporting general BT hyperboloidal beams. These techniques should take advantage from {\em global} GL expansions, as compared to {\em local} discretization schemes presently in use \cite{Shaugh2004,Vinet1992}.}

\item[{\em iii)}]{{\em Full} extension of the duality relations in \cite{Agresti2005a} to the family of BT hyperboloidal beams and supporting mirrors. Such an extension, demonstrated here for the dominant eigenmode, should be based on the ``complexification'' of the order of the involved Fourier transform operators. Finding such a one-to-one mapping between eigenmodes with {\em arbitrary} values of the twist-angle parameter would provide important physical insight and computational advantages.}

\end{itemize}

\begin{acknowledgments}
The work of J.A., E.D'A., and R.DS. is supported by the National Science Foundation under Grant No. PHY-0107417. The authors wish to thank Dr. R. O'Shaughnessy (Northwestern University, Evanston, IL) for useful comments and suggestions.
\end{acknowledgments}

\appendix
\section{Alternative Derivation of FM GL expansion coefficients}
\label{appendix}
It is instructive to illustrate the {\em direct} derivation of the GL expansion coefficients pertaining to the FM configuration, starting from the integral representation in (\ref{eq:U0}). In this framework, one first expands the modified Bessel function $I_0$ in series of Laguerre polynomials,
\beq
I_0\!\!\left(\frac{\sqrt{2}r_0 \xi}{w_0}\!\right)=\sum_{m=0}^{\infty} c_m L_m(\xi^2),
\label{eq:I0}
\eeq
where $\xi=\sqrt{2}r/w_0$, and
\begin{subequations}
\begin{eqnarray}
\!\!\!\!\!c_m\!\!&=\!\!&2\int_0^{\infty} \!\!L_m(\xi^2)\exp(-\xi^2)
I_0\!\!\left(\frac{\sqrt{2} r_0 \xi}{w_0}\!\right)\xi d\xi\label{eq:cm0}\\
&=&\frac{(-1)^m}{m!~2^{2m}}
\left(
\frac{\sqrt{2}r_0}{w_0}
\right)^{\!\!2m}\exp\left(\frac{r_0^2}{2w_0^2}\right).
\label{eq:cm}
\end{eqnarray}
\end{subequations}
Equation (\ref{eq:cm0}) follows from the Laguerre polynomials orthogonality condition \cite[Eq. 22.2.13]{Abramowitz}, whereas (\ref{eq:cm}) can be derived from \cite[Eq. (2.19.1.17)]{Prudnikov}. Substituting (\ref{eq:I0}) and (\ref{eq:cm}) into (\ref{eq:U0}) then yields, after some manipulation, the GL representation
\begin{eqnarray}
U_{0}(r,0)&=&\sum_{m=0}^{\infty}\!\psi_m(\xi) 
(-1)^m
\frac{\sqrt{2} w_0^2}{m! R_0^2}\nonumber\\
&\times&\int_{0}^{\frac{R_0^2}{2w_0^2}} \exp(-\zeta) \zeta^m d\zeta
\end{eqnarray}
whose expansion coefficients, recalling the integral expression of the incomplete Gamma function \cite[Eq. (6.5.1)]{Abramowitz}, coincide with those in (\ref{eq:map1}).

\bibliography{hyp_beam}

\end{document}